\documentclass[twocolumn,aps,prb]{revtex4}
\usepackage{graphicx}
\usepackage{amsmath}
\begin{document}

%\draft
\preprint{LA-UR-04-8345}
\title{Influence of supercoiling on the disruption of dsDNA}
\author{Shirish M. Chitanvis, Paul M. Welch}

\affiliation{
Theoretical Division,
 Los Alamos National Laboratory,\\
 Los Alamos, New Mexico 87545}
\date{\today}
\begin{abstract}
We propose that supercoiling energizes double-stranded DNA (dsDNA) so as to facilitate thermal fluctuations to an unzipped state.  We support this with a model of two elastic rods coupled via forces that represent base pair interactions. Supercoiling is shown to lead to a spatially localized higher energy state in a small region of dsDNA consisting of a few base pairs. This causes the distance between specific base pairs to be extended, enhancing the thermal probability for their disruption.  Our theory permits the development of an analogy between this unzipping transition and a second order phase transition, for which the possibility of a new set of critical exponents is identified.
\end{abstract}
\maketitle

\newpage
Unzipping of double-stranded DNA (dsDNA) as a prelude to transcription is a basic process of life\cite{topo2}.  There are presently two mechanisms that are considered for modeling this phenomenon.  One is the Peyrard-Bishop-Dauboix model, in which localization of energy via a  Fermi-Pasta-Ulam scenario of coupled non-linear oscillators causes thermal bubbles to form spontaneously at specific locations along a dsDNA chain\cite{kim}.  The other mechanism is modeled by representing proteins, which cause unzipping, as an external force\cite{nelson1,nelson2, bhatta}. This latter class of models envisions dsDNA as composed of coupled flexible Gaussian chains.  These models describe the unzipping transition as a second order phase transition. In reality\cite{bednar} the persistence length of dsDNA is approximately $45 \AA$.  It may be a reasonable approximation to treat a long semi-flexible chain as an effective Gaussian chain with the monomer length substituted by the persistence length in order to investigate global characteristics\cite{shirish0}. But a realistic attempt to treat the unzipping process through the application of an external force to a small region of the dsDNA must necessarily take into account the semi-flexible nature of the system.

There is an additional outstanding issue which complicates our understanding of the unzipping of dsDNA. It arises through the observation that one can estimate\cite{kim} the base-pair force as $\sim {\cal O} (100 pN)$, while the experimentally observed\cite{nelson1} minimum (external) disruptive force is only of the order of $\sim {\cal O}(10 pN)$.  Thus there would appear to be an intrinsic source of energy that allows the dsDNA to overcome this mismatch. What  subtle mechanism comes into play during base-pair separation to explain this mismatch?

Supercoiling is an ubiquitous feature of semiflexible rods\cite{siggia}.  One might conjecture that supercoiling plays a role in the packing of dsDNA.  But its role in the disruption of base-pair bonds has been paid only scant attention\cite{benham}.  In this Letter, we make the qualitative case that supercoiling energy can have an appreciable effect on this phenomenon, by providing extra energy to free energy barrier to disruption.

Benham\cite{benham,benham2} was the first to suggest a phenomenological model to describe the influence of supercoiling on base-pair disruption.  He obtained the parameters for his model by calibrating it to one set of dsDNA data, then verified that the same choice of parameters worked for another experiment.  Nelson et al.\cite{nelson1} also referred to supercoiling in passing. 
Hennig and Archilla\cite{hennig} have studied numerically an extended Peyrard-Bishop-Dauboix model to account for stiffness in dsDNA.  They focused numerically on the effect of an external force applied to dsDNA characterized by varying degrees of overtwist.  Their model is similar to the Poland-Scheraga model with torsion, which considers dsDNA as a semi-rigid one-dimensional ladder\cite{orland}.

The model developed in this Letter provides a deeper and more detailed, analytical insight into the mechanism of base pair disruption in dsDNA.
Marko and Siggia\cite{siggia} described the supercoiling of DNA by appealing to an evocative image of an over-twisted shoelace. The analogy that best describes the results of our model is that dsDNA is like a ladder made out of bamboo. If the ladder is twisted beyond its normal, equilibrium planar state, one can imagine that it would be easier to break the rungs.  We assert that this is precisely what happens in dsDNA.  Supercoiling, or over-twisting, energizes the dsDNA, making it more likely that thermal fluctuations will cause a given base-pair to be disrupted.  Our theory provides qualitiative insight into the experimental observations that superhelicity can make pBR322 DNA and {\em E. coli} duplex unwinding elements (DUE) susceptible to strand separation even if the over-twisting spans only a few base-pairs\cite{benham}.  These are presumed to be examples of generic strand separation behavior. Wang et al\cite{benham2} term this phenomenon stress induced duplex destabilization (SIDD).  Benham\cite{benham} and Wang et al\cite{benham2} provide a valuable tool by assuming an Ising-type model that accounts for superhelicity, and with which they can obtain results consistent with experimental observations.

We propose a model for dsDNA which is {\em ab initio} compared to Benham\cite{benham, benham2}.  It is based on generalizing the single elastica (conformations of a single, force-free elastic rod) model of Shi and Hearst\cite{shi} to coupled elastic strands.  Shi and Hearst begin with a Kirchoff rod, and manipulate the static equations of equilibrium into a non-linear Schrodinger equation!  Here the amplitude of the {\em wave-function} represents the curvature of the rod, while the argument of the phase factor is related to the torsion of the rod.  We then consider on this basis two elastic rods, each exerting an attractive force on the other at pre-set points along their backbones to represent base pairs. Our approach thus has the flavor of a Salpeter-Bethe equation for coupled quantum particles.  We derive a non-linear model analogous to the Kronig-Penney model in condensed matter physics to understand the effect of supercoiling on the coupling of the two coupled elastic rods.  The advantage of extending Shi and Hearst's formalism is that it allows us to use well-established methods for solving quantum mechanical problems to gain insight into the influence of supercoiling on dsDNA unzipping.

A straightforward description of our model follows through the prescription of an energy functional ${\cal E}$. When extremized with respect to its two component fields (each representing a strand), the functional yields a pair of non-linear Schrodinger equations (NLSEs), coupled via a (normal) force:

\begin{eqnarray}
{\cal E}&& = \int ds f(\psi(s))\nonumber\\
f(\psi(s)) &&= f^{0}(\psi_1(s)) + f^{0}(\psi_2(s)) - \Delta f(\psi_1(s),\psi_2(s)) \nonumber\\
{f^{0}} &&= \psi^*(s) \left( \frac{d^2}{ds^2} - c + \frac{1}{4} \vert\psi(s)\vert^2\right) \psi(s) \nonumber\\
\Delta {f} &&= 2 {\cal N}_{(2,1)}(s) \vert \psi_1(s) \vert + 2 {\cal N}_{(1,2)}(s) \vert \psi_2(s) \vert\nonumber\\
{\cal N}_{(2,1)}(s) &&= - {\cal N}_{(1,2)}(s)\nonumber\\
{\cal N}_{(2,1)}(s) &&= \Sigma_i {\sigma}_i ~ \delta(s-s_i)
\label{f1}
\end{eqnarray}
where the superscript zero denotes the functional associated with each elastica,
$\psi=\omega_x + i \omega_y$, with $\omega_x$ and $\omega_y$ being the components of the curvature vector in the $x$ and $y$ directions respectively\cite{shi}.  The phase of $\psi$ is related to the torsion $\tau$ of the strand\cite{shi}. 
We have considered a common label $s$ for the location along either chain.  This is sufficient to study the effect of super-coiling of a double-helix. We note that two independent labels would be required to address issues such as transcription, for which one strand is translated relative to the other. The generalization goes through in a straightforward fashion, but is an unnecessary complication for the current problem.

The subscripts $1,2$ refer to each of the two semiflexible strands in our dsDNA, $s$ refers to the location along the spine of the dsDNA; ${\cal N}_{(2,1)},{\cal N}_{(1,2)}$ refer to the attractive forces exerted by the strand labeled $2$ on the strand labeled $1$ and vice versa.  These forces are located at discrete positions along the backbone of the dsDNA, as signified by the Dirac delta function $\delta(x)$.  $\sigma_i$ represents the force due to the $i^{th}$ base-pair.  Using the potential developed by Rasmussen et al.\cite{kim}, we find that this force ranges from zero at the minimum of the potential well, and rises to a maximum of about $160 pN$ for AT (Adenosine-Thymine) base pairs at the inflection point of the potential before tailing off. The maximum force is approximately $417 pN$ for CG (Cytosine-Guanin) pairs.  The well-depths in these two cases are of the order of room temperature. 

As described by Shi and Hearst\cite{shi}, the parameter $c$ is a constant of integration and is related to the energies of bending and twisting.  
$c={\cal E}_b - Q^2/4$, where ${\cal E}_b$ is a dimensionless energy due to bending, and $Q$ is a dimensionless energy associated with torsion.
We shall treat it as an eigenvalue to be calculated to satisfy certain boundary conditions.
The energy ${\cal E}$ is chosen in units of $M~ \ell$, where $M \sim 2000 pN$ is the characteristic modulus of the dsDNA, and $\ell \sim 3.4 \AA$ is the average distance between base pairs.  The distances are measured with respect to $\ell$.

Equation \ref{f1} is based on a refined NLSE for an elastic rod (See Eqn (3.10) in the paper by Shi and Hearst\cite{shi}) in which certain transformations have been applied to cast the original relevant equations into a simpler form.  Moreover, we have retained only the component of the force between strands which is normal to each strand, while ignoring the tangential force. This provides an adequate model for coupled rods.

By casting the theory of Shi and Hearst\cite{shi} in terms of an energy functional ${\cal E}$, much physics can be gleaned in a manner similar to the Ginzburg-Landau theory\cite{binney}.  Consider only the polynomial part of the energy density for each strand. It can be written down schematically as follows (note that the curvature $\kappa=\pm \vert \psi \vert$):

\begin{equation}
f_{polynomial} = \frac{1}{4} \kappa^4 - c \kappa^2 - 2 {\cal N} \vert \kappa \vert
\label{fpoly}
\end{equation}

\begin{figure}[htbp]
%\begin{center}
%\includegraphics[angle=0,scale=.7]{dwell.pdf}
\includegraphics[angle=0,scale=.4]{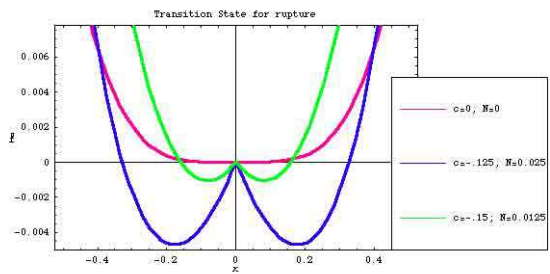}
\caption{{\bf Phase transition depicted schematically from a single-well to a double-well structure.}}
\label{fig1}
%\end{center}
\end{figure}

   Figure \ref{fig1} shows that for $c={\cal N} =0$, (red curve) there is a sole minimum at zero curvature ($\kappa=0$). This says, in essence, that when there is no binding force acting on the system and hence no bending or twisting, the minimum energy state is naturally a straight line.  This comes about due to the quartic, non-linear term in the expression which embodies the semiflexible nature of each strand.  This is a qualitative difference between the present case (red curve) and the ones described by the green and blue curves discussed in the next paragraph.
   
The green and blue curves in Fig. \ref{fig1} refer to configurations which possess different amounts of super-coiling. 
In either case, there are two minima, corresponding to two equal and opposite values of the curvature.  The attractive force dominates the behavior for small curvature, while the quartic term is important at high curvature. This results in the double-well structure. 
The blue curve refers to $c=-0.125$ and  ${\cal N} = 0.025$ (approximately $50 pN$).  
The green curve refers to $c=-0.15$ and  ${\cal N} = 0.0125$ (approximately $25 pN$).
The greater negative value of $c$ for the green curve indicates a relatively greater degree of super-coiling.
The state at the origin $\kappa=0$ can be thought of as a transition state (when the strand is linear).  The disruption of a base-pair corresponds to the achievement of this transition state. The well-depths, or equivalently, the barrier height between the wells dictates the probability that the system can reach the zero-curvature state through thermal fluctuations at a given temperature.  This is governed by the standard Boltzmann factor.  The green curve (with more negative c and therefore higher torsion) displays a lower barrier height as compared to the blue curve, and therefore is more susceptible to attaining the disrupted state ($\kappa=0$) via thermal fluctuations.

The qualitative change in the structure of $f_{polynomial}$ in the absence and presence of an attractive force is analogous to the critical point behavior in the theory of second order phase transitions, as the system passes from a single minimum to a double-well structure.  However, there is a major difference in the two theories, since the present case concerns an order parameter, viz., the curvature, which differs from the usual sort of order parameter (concentration, magnetization, etc.) which appears in Ginzburg-Landau theories.  Given that our model exhibits a transition reminiscent of a second order phase transition, one can obtain a certain amount of insight into the universality class it belongs to by appealing to a mean-field approximation.  In this case, by extremizing Equation \ref{fpoly}, one obtains a cubic equation for the order parameter $\kappa$.  One can examine its behavior as ${\cal N}\to 0$ and $c\to 0$, as it is the simultaneous vanishing of these two parameters which determines the transition point. 
In this sense each of these two parameters plays a role analogous to that of $b=T-T_c$, the temperature deviation from the critical temperature, in the standard Ginzburg-Landau theory.
If ${\cal N}\to 0$ faster than $c$, then one obtains the conventional exponent of $1/2$ since $\kappa ~\sim~ c^{1/2} \to 0$.  On the other hand, if $c \to 0$ faster than ${\cal N}$, then $\kappa ~\sim~ {\cal N}^{1/3} \to 0$ and suggests an exponent of $1/3$.  
Note that though there is but one independent variable in the model,  viz., the position along the backbone, one can obtain three-dimensional structures through the specification of the curvature and torsion of dsDNA.

Rudnick and Bruinsma have utilized the one-dimensional Poland-Scheraga model with torsion to suggest that there is a difference between a second order transition and denaturation in DNA\cite{rudnick}. 
Benham\cite{benham2} proposed an Ising-like model to describe base-pair disruption.  It should be possible to determine from his theory whether the unzipping transition can be described as a second order phase transition.  However, his energy functional is highly non-linear in the order parameter, and therefore difficult to analyze in a simple manner.  We speculate that it may be possible to map his model onto a non-linear {\em sigma} model (a continuum version of the Ising model) before analyzing it further.  This is beyond the scope of our paper. 

We mapped the conceptual idea of base-pair disruption discussed above directly to a double-helix.
By writing $\vert \psi(s)_p \vert = \sqrt{\psi^*_p(s)~\psi_p(s)}$,  $p=1,2$, one can extremize with respect to the complex conjugate fields $\psi^*_p$, and obtain a coupled set of NLSEs:

\begin{eqnarray}
\left( \frac{d^2}{ds^2} - c + \frac{1}{2} \vert\psi_p(s)\vert^2\right) \psi_p(s)&& - {\cal N}_{(p',p)}(s)~\frac{\psi_p(s)}{\vert \psi_p(s)\vert}=0\nonumber\\
&&~\forall p,p'=1,2;~p'\ne p
\label{f2}
\end{eqnarray}

Note that upon setting $\psi_2(s) \to 0$, the model reduces to that of Shi and Hearst\cite{shi} (see Eqn.(3.10)  in that paper).  The coupling between the rods occurs through the force exerted on one rod by the other.

As discussed in detail by Shi and Hearst, the amplitude of the wave-function is the curvature of the strand, and the argument of the phase factor is related to the torsion of the strand. As is well-known in differential geometry, curvature and torsion determine any given curve in space. Our solution of the NLSEs Eqn. \ref{f2} is based on the idea that the dsDNA is composed of two counter-twined helices.  According to the theorem of Bertrand\cite{sokolnikoff}, a helix is defined by the fact that the ratio of its curvature to its torsion is constant everywhere.  

With this in mind, we point out that a plane wave ansatz solves the NLSEs equation \ref{f2}, with ${\cal N}_{(2,1)}={\cal N}_{(1,2)}=0$ (the force-free case of uncoupled elastica) as follows:

\begin{eqnarray}
\psi_p(s) &&= \kappa_p(\tau_p)~\exp(i \tau_p s)\nonumber\\
\kappa_p(\tau_p) &&= \sqrt{ 2 (c + \tau_p^2)}
\label{f3}
\end{eqnarray}

$\kappa_p$ is the curvature, the total torsion in the strand is $\tau^{total}=  \tau_p + a/2$, 
$a = (2/\lambda-1) Q$, $\lambda$ is Poisson's ratio, and $Q$ is a measure of the torsional energy\cite{shi}.
We see in this case that the ratio of the curvature to the torsion is a constant everywhere.
This occurs because $\tau_p$ and, therefore, the amplitude of the {\em wave-function} is independent of the variable $s$. As such we observe that a plane wave solution to our force-free NLSE is consistent with a helix.

We envision the solution of the full NLSEs with non-zero attractive forces between them by connecting successive helical solutions across each delta-function force. This is achieved by integrating the NLSE for each strand over an infinitesimal interval across any given delta-function force.  This yields a condition connecting the first derivative of $\psi$ on either side of the delta-function force. 
We thus get a complex transcendental equation which yields two conditions (real and imaginary parts of the condition considered separately) to connect the amplitude and phase of $\psi$ on either side of the delta-function.  These conditions can be solved in a recursive manner to get values for the torsion and the parameter $c$ in successive segments of the chain.  If the phase difference between the solutions on either side of the delta-function force is given by $\delta_n$ at the $n^{th}$ node, then the solution is given by the following set of recursive relations:

\begin{eqnarray}
\tan\delta_n &&=\frac{\sigma_n}{\tau_{n-1} ~\kappa_{n-1}} \nonumber\\
\kappa_{n} &&= \sqrt{2 (c_n + \tau_n^2)} \nonumber\\
c_n &&= \frac{\tau_{n1}^2 (c_{n-1} + \tau_{n-1}^2) \sec(\delta_{n})^2} {(\tau_{n-1}+\delta_n)^2} - (\tau_{n-1}+\delta_n) \nonumber\\
\tau_n &&= \tau_{n-1} + \delta_{n} 
\label{recur}
\end{eqnarray}
Knowing the parameters for the helix at one end of the chain as a boundary condition, one can march recursively down each strand.
In this sense we have solved an initial-value problem.
Given that the dsDNA is two counter-twined helices ($\kappa \to -\kappa$), the parameters for the other helix are immediately obtained.

The cartesian co-ordinates of a helix representing each segment between two consecutive delta-function forces are given parametrically as\cite{sokolnikoff}: $x_n(\theta)= x_n^{(0)}+a_n \cos(\theta)$, $y_n(\theta)=y_n^{(0)}+ a_n \sin(\theta)$, $z_n(\theta) = y_n^{(0)}+ k_n \theta$, where $\theta$ is the azimuthal angle.  
$x_n^{(0)},y_n^{(0)},y_n^{(0)}$ serve to fix the starting point of the $n^{th}$ segment. 
The parameters $a_n$ and $k_n$ are related to the curvature and torsion of the helix.
$k_n$ is the pitch of the helix, and $a_n$ is the radius of the cylinder around which the curve is wound.  These parameters are related to the curvature and torsion as follows: 

\begin{eqnarray}
a_n &&= \frac{\kappa_n}{ \kappa_n^2 + (\tau^{total}_n)^2 }\nonumber\\
k_n &&= \frac{\tau^{total}_n}{\kappa_n}
\label{hel1}
\end{eqnarray}

Armed with this formal solution, we now display two cases to illustrate the notion that adding extra twist to dsDNA facilitates the path to the disruption of base-pairs. 
The parameters chosen for our calculations are representative of dsDNA, but do not refer to any specific DNA sequence. This was done to display the conceptual basis of the model and provide a base-line for more realistic calculations which must follow.

\begin{figure}[htbp]
%\begin{center}
\includegraphics[angle=0,scale=.25]{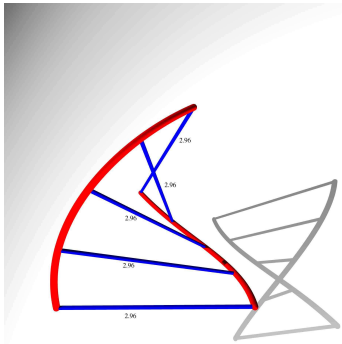}
\caption{{\bf Double-helical structure displayed over five sections for $c_1=-0.15,~\tau_1=0.425, a=0.2$. The distance between base-pairs is marked on each rung, in units of $3.4\AA$.
We have approximately ten base-pairs per turn for this case and it represents the ground state of dsDNA.
There is a light source at the top left, casting a shadow of the double-helix.}}
\label{fig2}
%\end{center}
\end{figure}

 In the first case (see Fig. \ref{fig2}), the constants
(boundary conditions) are $c=-0.15$, $a=0.2$ and $\tau_p=0.425$ for the first segment.  This yields approximately $10$ base-pairs per turn.  The base-pair separation is approximately $10\AA$ as in normal DNA.
Hence it was designated the ground state of a model dsDNA and the forces between the base-pairs were set to zero.  This would be consistent with an effective potential representing base-pairs\cite{kim} which has a minimum at the equilibrium configuration, and therefore a null force.

\begin{figure}[htbp]
%\begin{center}
\includegraphics[angle=0,scale=.25]{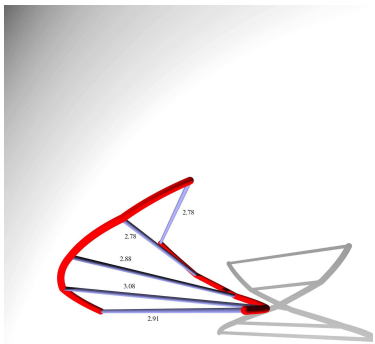}
\caption{{\bf Double-helical structure displayed over five sections for $c_1=-0.15,~\tau_1=0.46, a=0.2$.
There is more torsion in this case as compared to the previous figure.
Base-pairs are separated increasingly compared to Fig.\ref{fig2}, but in a {\em localized} fashion,
like the thermal bubbles calculated by Rasmussen et al.\cite{kim}.
The use of delta-function forces in the model permits kinks in the structure.
}}
\label{fig3}
%\end{center}
\end{figure}

In the second case (Fig. \ref{fig3}) we took $\tau_p=0.46$, keeping the other boundary constants unchanged. Thus this is an over-twisted case. 
This causes the double-helix to be squeezed down such that the distance between the base-pairs increases as we proceed up the ladder. The force was taken to be approximately $20 pN$.
Our estimate for the force is in qualitative concert with the potential of Kim et al\cite{kim}.
Unzipping, or base-pair disruption in our model, is signaled by a tendency of the base-pairs to show increasing separation.  
It is important to note, however, that the increase in separation, indicated in units of $\ell=3.4\AA$ in the figure along each base-pair, is not monotonic.  
as this happens, the absolute value of $c_n < 0$ increases in successive segments, signaling a lowering of the free energy barrier depicted schematically in Fig.\ref{fig1}.
There is a clear tendency of the separation to increase, followed by a decrease.
This indicates that supercoiling is not distributed uniformly, but rather, can be localized over a small region of dsDNA.  This is consistent with the model of Rasmussen et al. in which vibrations get localized to generate thermal bubbles\cite{kim}.  Indeed it is possible that supercoiling and vibrations and supercoiling act in concert to achieve base pair disruption.

There exist examples of genes that are only expressed once the relevant section of DNA is in a super-coiled state\cite{benham}.  It should be possible within the framework of our model to predict whether such a sequence is an initiation site for transcription.
At the moment we can explain qualitatively via this illustrative example, the experimentally known fact that in pBR332 DNA and {\em E. coli}\cite{benham}, supercoiling enhances unzipping.  The examples just discussed show that given a certain torsion and curvature, progression over a finite number of base-pairs can create a localized increase in base-pair separation.  This will lead to a lower attractive force. Eventually, the probability of a thermal fluctuation to a state of zero curvature (disruption) can occur with appreciable frequency, as conceptualized in Fig.\ref{fig1}.

We have restricted attention for the moment to the description of only a short sequence of like base-pairs, as that is sufficient to display the localization of super-coiling.  We are solving an initial-value problem in which the conditions at one end of the double-chain are utilized to march recursively down the sequence.  Therefore adding an extension to any given sequence will not affect the preceding section of dsDNA.

In summary, we have extended the single-strand theory of Shi and Hearst\cite{shi} to two interacting elastic rods connected to each other via attractive forces.  Their language is thus developed to the point where we can provide not only an over-arching framework for the unzipping transition as analogous to a second-order phase transitions, but also puts us in a position to predict the behavior of specific sequences in future work.  The analytical solution provided in this paper can be generalized to more complex forces acting between the strands by using numerical techniques. This can yield predictions for specific sequences occurring in a given dsDNA which can be manipulated experimentally to change the supercoiling.

We would like to thank Kim Rasmussen and Tommy Sewell for a critical reading of the manuscript.  The work was performed under an LDRD project at Los Alamos National Laboratory.


\begin{references}
\bibitem{topo2} Wilsterman, A.M. and Neil Osheroff, Curr. Topics Med. Chem. {\bf 3}, 201 (2003).
\bibitem{kim} G. Kalosakas, K. ¯. Rasmussen, A. R. Bishop,
C. H. Choi and A. Usheva, Europhys. Lett. {\bf 68} , 127 (2004).
\bibitem{nelson1} Lubensky, D.K., Nelson, D.R., Phys. Rev. E {\bf 65}, 031917 (2002).
\bibitem{nelson2} Weeks, J.D., Lucks, J.B., Kafri, Y., Danilowicz, C., Nelson, D.R., Prentiss, M.,
xxx-e-archive cond-mat/04056246, (2004).
\bibitem{bhatta} Bhattacharjee, S.M., J. Phys. A {\bf 33}, L423 (2002). 
\bibitem{bednar} Bednar, J., Furrer, P., Katritch, V., Stasiak, A.Z., Dubochet, J. and Stasiak, A., J. Mol. Bio. {\bf 254}, 579 (1995).
\bibitem{shirish0} S.M. Chitanvis, Phys. Rev. E {\bf 63}, 021509 (2001).
\bibitem{hennig} D. Hennig and J.F.R. Archilla, Phys. A {\bf 331}, 579 (2004).
\bibitem{orland}  Garel, T., Orland, H., and Yeramian, E. xxx-e-archive q-bio.BM/0407036 (2004).
\bibitem{siggia} Marko, J.F. and  Siggia, E.D., Phys. Rev. E {\bf 52}, 2912 (1995).
\bibitem{benham} Benham, C.J., J. Mol. Bio. {\bf 225}, 835 (1992).
\bibitem{benham2} Wang, H.Q. , Noordewier, M , Benham, C.J., Gen. res. {\bf 14}, 1575 (2004).
\bibitem{shi} Shi, Y., and Hearst, J.E., J. Chem. Phys. {\bf 101}, 5186 (1994).
\bibitem{binney}Binney, J.J., Dowrick, N.J., Fisher, A.J. and Newman, M.E.J, {\em The Theory of Critical Phenomena}, Clarendon Press, Oxford (1992).
\bibitem{rudnick} Rudnick, J., Bruinsma, R., Phys. Rev. E {\bf 65}, 030902(R) (2002). 
\bibitem{sokolnikoff} Sokolnikoff, I.S., {\em Tensor Analysis: Theory and Applications to Geometry and Mechanics of Continua}, 
John Wiley and Sons, New York (1951).
\end{references}
\end{document}